# Study of F4TCNQ dopant diffusion using transport measurements in organic semiconductors

Meghan P. Patankar, Kapil Joshi[a] and K.L.Narasimhan[*b]

Dept. Of Condensed Matter and Materials Science, Tata Institute of Fundamental Research, Homi Bhabha Road, Colaba, Mumbai 400005.

Abstract:

In this paper, we report on electrical transport in F4TCNQ doped organic semiconductor (host) materials. By monitoring the conductance of the sample in-situ during and after deposition, we show that sequential deposition of F4TCNQ and host semiconductor results in bulk doping of the semiconductor. In addition, the doping density (and conductivity) of the host can be easily controlled by adjusting the thickness of the bottom F4TCNQ layer. This alternative scheme for doping is simpler than the conventional way of doping small molecules which involves simultaneous co-evaporation of host and dopant. In the sequential doping scheme outlined here, bulk doping of the host takes place due to rapid diffusion of F4TCNQ in the host material. The motion of F4TCNQ in the host is complex and is not always described by a simple diffusion process. In situ transport measurements provide a quick and easy way of measuring dopant diffusion in new hosts. Based on the doping results, we also outline a possible route to improved transconductance for organic thin film transistors in a manufacturing environment.

Keyords: Organic semiconductors, doping , electrical transport, F4TCNQ, contacts a Present address  Labindia Instruments Pvt. Ltd., L.B.S.Marg, Thane 400062 India

b Present address- EE, IIT Bombay, Powai Mumbai 400076

∗ corresponding author

email:kln@mailhost.tifr.res.in

Introduction:

In recent years there has been tremendous progress in light emitting devices, field effect transistors, solar cells etc. based on organic semiconductors [1]. Electrical contacts play an important role in facilitating charge injection in these devices. Doping of semiconductors makes it possible to reduce contact resistance. In organic semiconductors, doping relies on charge transfer from the host to the dopant molecule [2-8]. A widely used p type dopant is F4TCNQ [2-9]. This molecule has a deep LUMO level (-5.2 eV) which is energetically in the vicinity of the HOMO level of many organic semiconductors [4]. Doping is facilitated by charge transfer from the HOMO level of the host to the LUMO of the dopant molecule. Doped molecular layers are formed by co-evaporation of the dopant molecule (guest) with the molecule of interest (host) [2-7,9]. For polymers, doping has been demonstrated by mixing F4TCNQ with the polymer before spin coating of the polymer [8]. The use of doped contact layers has resulted in low operating voltages in light emitting diodes [2,3].

F4TCNQ is known to be a fast diffusing species in ZnPc and CBP (4,9). These results have been established by photoemission studies. In this paper, we use the rapid diffusion of F4TCNQ to create bulk doped layers by sequential deposition of dopant (F4TCNQ) and organic semiconductor (host) as opposed to co-evaporation of the dopant and the host. This is a much simpler alternative and has the advantage that it can also be easily implemented in a roll to roll manufacturing processes. Measurement of the conductance of the sequentially deposited dopant layer (F4TCNQ) and organic semiconductor (host) layer in situ during and after deposition of the different layers allows a study of the diffusion process on a significantly smaller time scale than is possible in a photoemission experiment. The high sensitivity and ease with which these experiments can be performed is an additional benefit and provided the motivation for this work.

Experiment:

The samples reported here are made by vacuum deposition unless specified otherwise. The glass substrates had ITO contact pads separated by a 1 mm gap to facilitate planar conductance measurements. All depositions are carried out at a pressure of $8.10^{-7} - 10^{-6}$ Torr pressure. The typical deposition rate for the organic layers is of 0.1- 0.2 nm/sec. In this paper, all the layers are deposited sequentially one after another and not co-evaporated. The deposition sequence consists of a F4TCNQ layer followed by the deposition of the organic semiconductor (TYPE A samples) or conversely the inverted structure where the organic semiconductor layer is first deposited followed by F4TCNQ on top (Type B samples). Both Type A and Type B samples are deposited in the same pumpdown using a shutter. Type B samples, prior to the deposition of the top F4TCNQ layer, also serves to provide the reference conductance of the undoped semiconductor layer. Unless otherwise mentioned, the thickness of the F4TCNQ layer and the organic semiconductor layer is 3 nm and 40 nm respectively. The temperature of the sample holder could be varied from 180K to 450K. Most of the depositions were carried out at 300K. In some cases, deposition was also carried out at 180K.

The organic semiconductors investigated are TPD, α-NPB, Pentacene, Rubrene and P3HT. In the case of P3HT samples, a 100 nm thick P3HT film was deposited by spin coating on the ITO substrates and then transferred to the vacuum system for F4TCNQ deposition. Hence only type B samples were possible for P3HT films.

All the samples are measured in a planar configuration in situ during and after deposition of the organic layers in situ in the deposition system. The conductance measurements are made at low electric field (50-150 V/cm). The current- voltage characteristics is verified to be Ohmic. The choice of ITO as electrode is governed by the fact that the samples that are being investigated are hole conductors. In addition to the conductance measurements at 300K, in some cases, we also measured, in situ, the temperature dependence of the conductance of the samples between 200-400K immediately after deposition..

Results and Discussion:

Figure 1 shows the thickness of a sequentially deposited F4TCNQ and Pentacene (TYPE A) layers at 300K as a function of deposition time. The figure also shows the conductance of the sample monitored during and after deposition of the layer. The planar conductance of the sample is $\sim 10^{-12}$ $\Omega^{-1}$ after the deposition of a 3 nm F4TCNQ layer. After a few minutes pentacene is deposited at 0.2 nm/s. The conductance of the sample increases continuously during deposition (as the thickness of the pentacene layer increases) and reaches a constant value when the pentacene deposition is stopped. The conductance of the sample increases over four orders of magnitude. In another experiment, the F4TCNQ layer was first deposited at room temperature. The sample was then cooled to 180K followed by deposition of pentacene. Figure 1 also shows the conductance of the F4TCNQ/pentacene layer at 180K. The conductance at 180 K again increases with pentacene layer thickness even though pentacene is deposited at low temperature.

Figure 2 shows the conductance of a F4TCNQ/TPD layer (type A sample) during deposition of the TPD layer for two different thickness of the bottom F4TCNQ layer (3nm and 30 nm). The conductance increases with the TPD thickness for both samples. The conductance of the sample with 30 nm F4TCNQ layer thickness has a conductance seven times larger than the sample with a 3 nm bottom F4TCNQ layer. Similar results were also obtained for Pentacene samples. The conductance of the F4TCNQ/TPD layer is much smaller than the F4TCNQ/Pentacene layer for the same F4TCNQ bottom layer thickness. Results similar to that for TPD were also obtained for α-NPB films (not shown).

Figure 3 shows the conductance of F4TCNQ/Rubrene for a type A sample. The conductance of rubrene is not increased by the presence of the bottom F4TCNQ layer. The figure also shows the conductance when TPD is deposited on top of F4TCNQ/Rubrene layer. The conductance of the F4TCNQ/Rubrene/TPD layer increases with the thickness of the TPD layer.

We now discuss type B samples. In this experiment a the organic semiconductor layer is first deposited followed by the F4TCNQ layer. Figure 4 shows the in situ conductance (monitored during growth ) of a 40 nm thick pentacene layer deposited at 300K. The conductance of the pentacene layer (in the absence of the F4TCNQ layer) is very small ($<10^{-12}$ $\Omega^{-1}$). Figure 2 also shows the large change in the conductance ( about four orders of magnitude) on evaporation of a thin (3nm) layer of F4TCNQ on top of the Pentacene layer. When the deposition of the F4TCNQ layer stops, the conductance drops by about a factor of two and is then constant.

The conductance of P3HT (Type B) is very sensitive to the presence of the top F4TCNQ layer. For less than a monolayer deposition of F4TCNQ (as measured by a quartz crystal monitor), the conductance of P3HT increases by five orders of magnitude. The temperature dependence of conductance of the doped film is also very small suggesting that these samples are degenerately doped. These results show that P3HT can be heavily doped by deposition of just a monolayer of F4TCNQ on top of a precast P3HT film.

Type B samples of TPD and α-NPB exhibit a very different behavior. In these samples, the conductance remains very small when F4TCNQ is deposited on top of the semiconductor – the top F4TCNQ layer does not dope type B samples of TPD and α-NPB. We now discuss these results.

With the exception of rubrene, we see that for all the type A samples (organic semiconductor on F4TCNQ), the conductance of the organic semiconductor increases significantly with the thickness of the semiconductor layer and plateaus off when the deposition of the organic semiconductor stops. The conductance of the bare organic semiconductor ( and bare F4TCNQ layer) is typically $<10^{-12}\ \Omega^{-1}$ - much smaller than the conductance of the bilayer viz. F4TCNQ/organic semiconductor samples. This suggests that F4TCNQ dopes the organic semiconductor. This is understandable as the LUMO of F4TCNQ is very deep (- 5.2 ev) and acts as an efficient hole dopant in many systems. Since the conductance increases with the thickness of the organic semiconductor layer (Type A samples), we conclude that F4TCNQ diffuses rapidly through the semiconductor layer doping the bulk of the material giving rise to the large increase in conductance. Rapid diffusion of F4TCNQ has been demonstrated in photoemission experiments in CBP and ZnPc [4,9]. In these experiments, the authors demonstrated that F4TCNQ diffused through 10 nm of sample in about ten minutes – the time taken to move the sample from the preparation chamber to the measurement chamber. The experiments reported here allows us to monitor the diffusion using conductance as a probe with higher sensitivity and on much smaller time scales and also measure the diffusion at low temperature. Quite surprisingly, F4TCNQ has a large diffusion coefficient even at 180K (as measured by the conductance change in pentacene in fig.1).

For some systems (TPD,NPB) the conductance change for type A samples (F4TCNQ/semiconductor) is very different from that for type B samples (semiconductor/F4TCNQ). The deposition of F4TCNQ on top of TPD does not cause any change in the conductance of the TPD/F4TCNQ layer. This implies that F4TCNQ does not diffuse down through the TPD layer towards the substrate side (for type B samples). The motion of F4TCNQ in the semiconductor hence cannot be described by a classical diffusion process. An analog of this exists in polymers. In many organic polymer blends made by spin coating, vertical segregation of the blends has been reported [10,11]. The segregation is believed to arise due to the different free energy at the polymer-substrate interface and polymer-air interface. The solvent also plays an important role in the vertical segregation. It is interesting that even in evaporated samples (like the ones reported here) vertical segregation appears to take place. It appears that for the F4TCNQ/TPD system, the F4TCNQ prefers to be at the semiconductor-air interface rather than the substrate-semiconductor interface. We have also observed similar results for α-NPB:F4TCNQ.

In the case of rubrene, the presence of the bottom F4TCNQ layer does not significantly change the conductance of rubrene. We hence conclude that F4TCNQ does not dope rubrene. Since the conductance of the sample increases when a TPD layer is deposited on rubrene, we conclude that F4TCNQ diffuses through rubrene and on deposition of a TPD layer diffuse through the TPD layer, doping the TPD layer - giving rise to the observed conductance in fig.4. The conductance change for different host materials deposited on a F4TCNQ layer of the same thickness is different for different host materials. We now try to qualitatively understand the different relative doping efficiency of F4TCNQ for the different systems described here based on a simple model for doping.

Doping of organic semiconductors can be understood based on a simple model of disorder. The electronic structure of organic semiconductors can be described by a HOMO and LUMO level – both of which are broadened due to disorder. This is shown schematically in Figure 5. The HOMO and LUMO level broadening is assumed to be a gaussian distribution centred around HOMO and LUMO levels and can be described as [12,13]

$$g(E) = g_0 \exp\{-[(E - E_{HOMO})/\gamma]^2\} \quad \dots\dots\dots\dots\dots\dots\dots\dots\dots\dots\dots\dots 1$$

where $\gamma$ is the Gaussian broadening parameter, $E_{HOMO}$ the peak position of the Gaussian and $g_0 = N/\sqrt{(2\pi\gamma)}$ where N is the total number of states. Levels above $E_{HOMO}$ (below $E_{LUMO}$) act as deep trap levels. For a pure semiconductor ( no impurities), the position of the fermi level will lie at the energy where the tail of the HOMO and LUMO levels overlap. In fig.5a, the fermi level is shown to be in the middle of the band gap – $E_g/2$. Hole doping of an organic semiconductor requires the LUMO of the dopant molecule to be energetically close to the HOMO of the host. The LUMO of F4TCNQ has been determined to be at -5.2 eV with respect to vacuum from photoemission experiments. This is deep enough to lie close to the HOMO of many host molecules. Doping involves charge transfer from the HOMO of the host to the LUMO of the dopant. This is shown schematically in figure 5b. For a broad density of states, charge transfer takes place from the deepest HOMO of the host to the LUMO of the dopant. Depending on the dopant concentration and the density of deep (HOMO) traps, the fermi level will move towards the dopant LUMO level. This is indicated schematically in figure 5b. The Fermi level position and the electrical activation energy ($\Delta E$) is also indicated in the figure. Doped organic semiconductors are closely compensated semiconductors. The doping efficiency depends on the gaussian broadening parameter $\gamma$, the dopant density and the HOMO (host)- LUMO (dopant) separation ($\delta$). The value of $\gamma$ influences transport as follows:

a) The effective transport energy ($E_t$) can be written as $\gamma^2/kT$ where k is the Boltzmann constant. The effective transport level moves towards the peak of the HOMO distribution as $\gamma$ decreases [14]. This will result in higher mobility.

b) The doping efficiency also depends on the magnitude of $\gamma$. For a broad Gaussian (large $\gamma$), the doping efficiency is much lower than for a small value of $\gamma$.

The HOMO levels for the molecules investigated here lie between 5.2- 5.6 eV. However, the value of $\gamma$ can be very different for these systems. The large conductance increase for P3HT for a very small dopant concentration indicates that both $\gamma$ and $\delta$ are small for P3HT. This accounts for the large doping efficiency seen for F4TCNQ doping of P3HT. This result is consistent with the fact that in hole-only structures, electrical transport in P3HT is space charge limited with the current proportional to $V^2$ where V is the voltage across the device [15]. This implies that the deep level density is small - and consistent with a small value for $\gamma$.

In the case of pentacene, for the same bottom layer thickness of F4TCNQ (30 nm), the increase in conductance for a F4TCNQ/Pentacene layer is very much greater than for a F4TCNQ/TPD layer. We have measured the temperature dependence of conductivity for the two samples. The electrical conductivity of the two samples is thermally activated. The activation energy is different for the two samples viz. 0.27 eV for F4TCNQ/TPD and 0.08 eV for the F4TCNQ/Pentacene layer. The smaller electrical activation energy seen in F4TCNQ/ Pentacene films can be explained as follows:

a) If the Gaussian broadening parameter $\gamma$ in eq. 1 is smaller in pentacene than in TPD, then the doping efficiency of F4TCNQ is larger. Since the hole mobility in pentacene (0.1-1 $cm^2$/ V-s) is larger than the hole mobility in TPD ($10^{-2} – 10^{-3}$ $cm^2$/Vs), this is a very likely possibility.

b) In addition, if $E_{HOMO}$ for Pentacene lies above $E_{HOMO}$ for TPD, this will also contribute to a smaller activation energy.

We now compare the electrical activation energy of a doped TPD film made by co-evaporation of F4TCNQ and TPD (4% F4TCNQ:TPD co-evaporated film) with a sequentially deposited F4TCNQ (30 nm)/TPD film. The electrical activation energy for the co-evaporated film is 0.38 eV [16] as against 0.27 eV for the samples made by sequential deposition. This shows that doping in sequentially deposited F4TCNQ/organic semiconductor structures is comparable to doping in co-evaporated films. The doping level can be controlled very simply by adjusting the thickness of the bottom F4TCNQ layer in the sequentially deposited film.. This hence provides greater control with the advantage of simplicity.

The fact that the conductivity of Rubrene does not increase when deposited on F4TCNQ is presumably due to the fact that the HOMO level in Rubrene is deeper than in Pentacene. F4TCNQ diffuses through the Rubrene layer very rapidly during Rubrene deposition as is evidenced by the fact that deposition of TPD on top of Rubrene contributes to an increase in the conductance (by doping the TPD). It is tempting to attribute that the observed rapid diffusion of F4TCNQ through the organic layer is driven by the heat of deposition when the organic layer is deposited on F4TCNQ. However, we see from fig.1 that for pentacene films deposited at low temperature (180K) the diffusion of F4TCNQ is quite rapid. The mechanism of the rapid diffusion of F4TCNQ is not clear.

Doped hole injection layers play an important role in reducing contact barriers in devices. These experiments show that for F4TCNQ doping, co-doping is not necessary- the doped layers can be obtained by sequential deposition of F4TCNQ followed by the deposition of the hole transport layer. Since the conductance change depends on the thickness of the bottom F4TCNQ layer (fig.2), this implies that the doping concentration in the organic layer can be controlled by the thickness of the F4TCNQ layer. In many of our devices, we have routinely used sequentially deposited F4TCNQ/TPD layers as good hole injection layers [17] instead of co-evaporated F4TCNQ:TPD layers.

We also explored the possibility of using doped P3HT as a conducting anode and HIL in ITO/P3HT/F4TCNQ/Alq3/LiF/Al sandwich devices. Figure 6 shows the current –voltage characteristics of this device. The figure also shows the current voltage characteristics of ITO/F4TCNQ(3nm)/TPD/Alq$_3$/LiF/Al device for comparison. We see from the figure that while the injection in the doped P3HT anode sample is less efficient than that in the F4TCNQ/TPD sample, the injection levels are reasonable.

The results reported here have important implications for printed organic thin film transistors using P3HT or pentacene as the active layer. The series resistance arising from the contact resistance of the source (drain) contacts can be reduced by first printing a F4TCNQ layer before printing the source and drain contacts. The diffusion of F4TCNQ through P3HT (pentacene) will ensure low contact resistance regions for source and drain – improving FET response and enable short channel transistors which are currently constrained by the source (drain) contact resistance.

In conclusion, we have shown that it is possible to make highly doped layers by a sequential deposition of a thin layer (3-30 nm) of F4TCNQ followed by the organic semiconductor. This is a much simpler alternative to co-evaporation of the dopant and host layers. In addition, we demonstrate that the amount of doping can be easily controlled by adjusting the thickness of the F4TCNQ layer. The doping occurs due to rapid diffusion of F4TCNQ through the host layer as the host layer thickness is being built up. The diffusion of F4TCNQ through the host can be easily studied using in-situ conductance measurements. This technique provides an efficient and simple way for rapid evaluation of the diffusion of new dopants in different host materials. The motion of F4TCNQ through the host is not a simple diffusion process – for TPD and α-NPB host material, diffusion of F4TCNQ through the host is inhibited if F4TCNQ is deposited on top of TPD- in contrast to results on P3HT and Pentacene. The doping efficiency depends on the width of the Gaussian that characterizes the semiconductor density of states and the energy difference between the LUMO of F4TCNQ and the HOMO of the semiconductor. A knowledge of the doping efficiency enables us to get a qualitative idea of the magnitude of the disorder. These experiments have important implications for technology of printed transistors. For top source – drain P3HT (Pentacene) transistors made by printing, a F4TCNQ layer printed prior to the deposition of source and drain contacts will dope the contact regions. This will facilitate a significant lowering of the contact resistance improving the transconductance of organic thin film transistors and enable fabrication of short channel transistors.

Acknowledgments: We thank Professor N.Periasamy for useful discussions.

Figure captions:

1. Figure 1: A semi log plot of the conductance of a F4TCNQ/ pentacene film vs time (during growth) grown at two different temperatures ○ 300K and *180K. Open symbols represent conductance and filled symbols the thickness. The pentacene film thickness vs time is shown to show the evolution of the conductance with pentacene thickness. The F4TCNQ layer was deposited at 300K.

2. Figure 2: A semilog plot of the conductance of a F4TCNQ/TPD film for two different F4TCNQ thickness * 30 nm and ○3 nm vs TPD thickness (during growth) of the TPD film. The thickness (growth) of the TPD film as a function of time (●) is also shown to correlate the conductance rise with the TPD thickness.

3. Figure 3: ● A plot of the thickness as a function of time (during growth) for a F4TCNQ/Rubrene/TPD layer. The F4TCNQ layer is not shown and has a thickness of 3 nm. The arrows indicate the beginning and end of the Rubrene (TPD) layer as a guide to the eye. □ A plot of the conductance of the growing layer during the deposition process.

4. Figure 4: ● A plot of the thickness (during growth) of a type B sample of Pentacene (30nm) followed by F4TCNQ(3nm) film as a function of time. The arrow on top marks the beginning of the F4TCNQ deposition. □ A plot of the conductance of the pentacene/F4TCNQ layer as a function of time (during growth of the film). The conductance increases sharply when the F4TCNQ deposition begins.

5. Figure 5: The density of states and the position of the Fermi level in a) an undoped sample b) doped sample. The red line represents the LUMO level of F4TCNQ. The conductivity activation energy ΔE and the shift in the Fermi level due to doping $\Delta E_f$ is also indicated in the figure. See text for details.

6. Figure 6: A plot of the current density vs Voltage for
   (Δ) ITO/F4TCNQ/TPD/Alq$_3$/BCP/LiF/Al and
   (■) ITO/P3HT/F4TCNQ/Alq$_3$/BCP/LiF/Al devices.

Figure 1

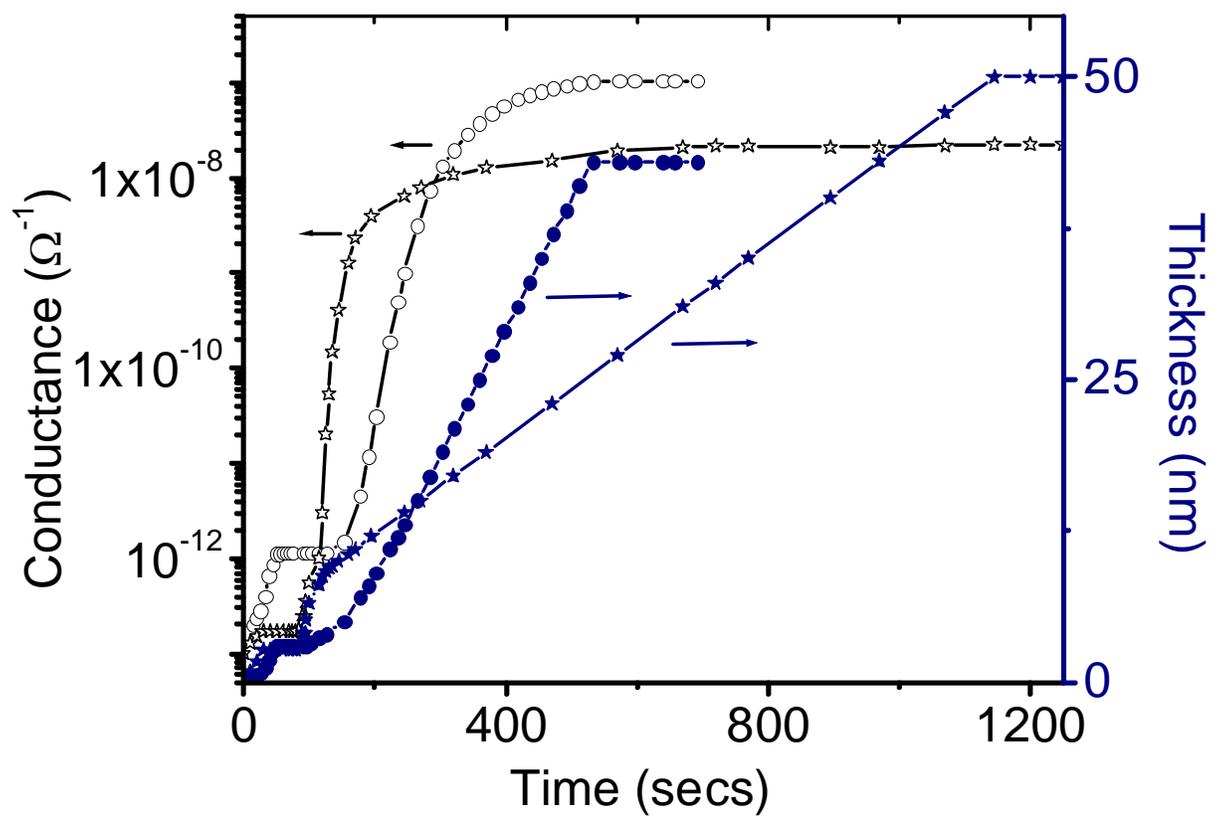

Figure 2

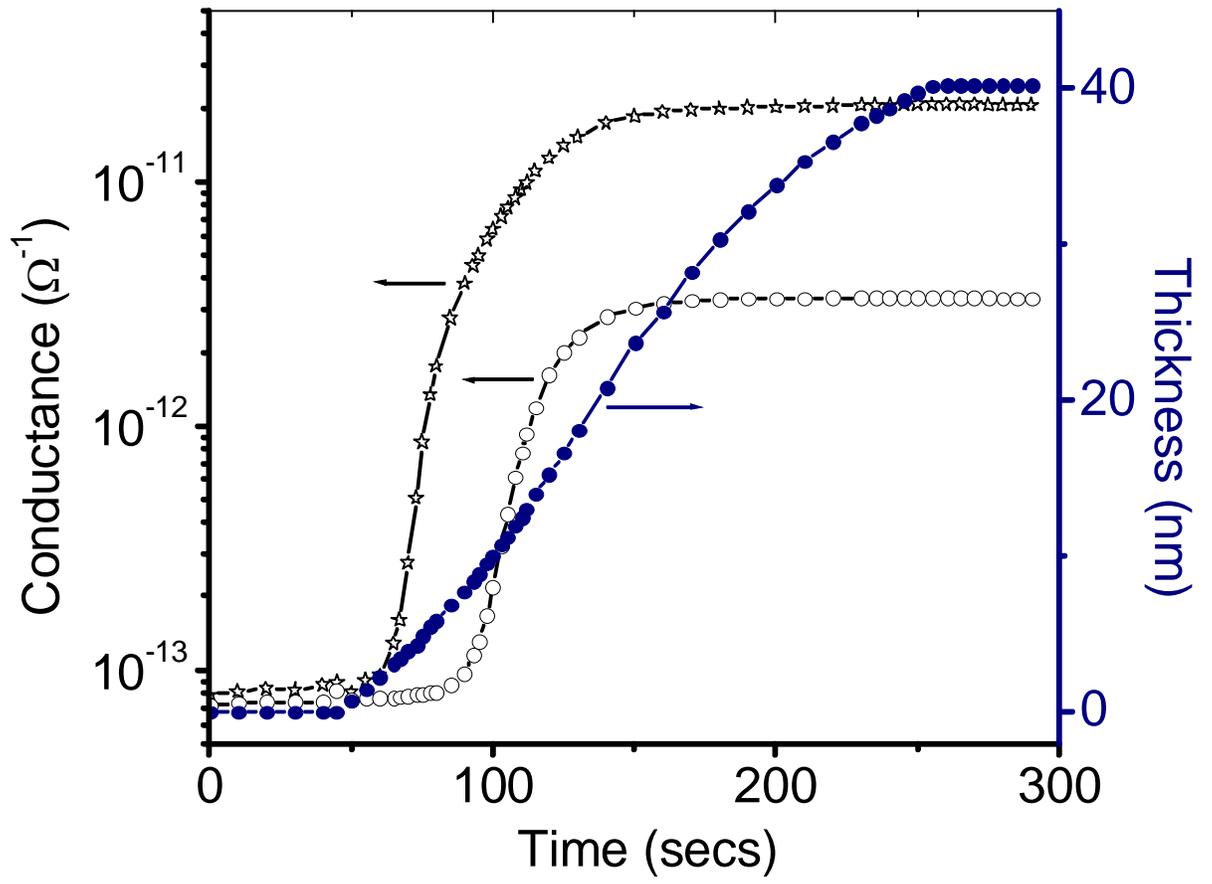

Figure 3

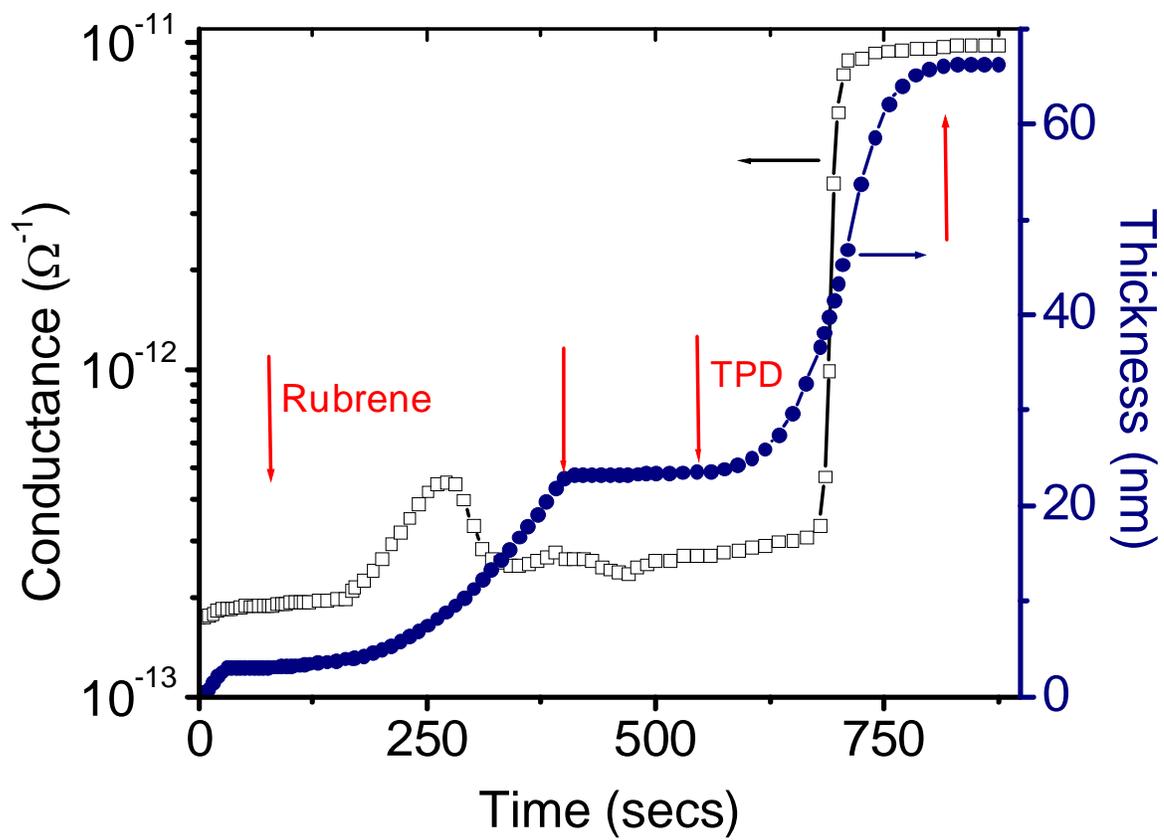

Figure 4

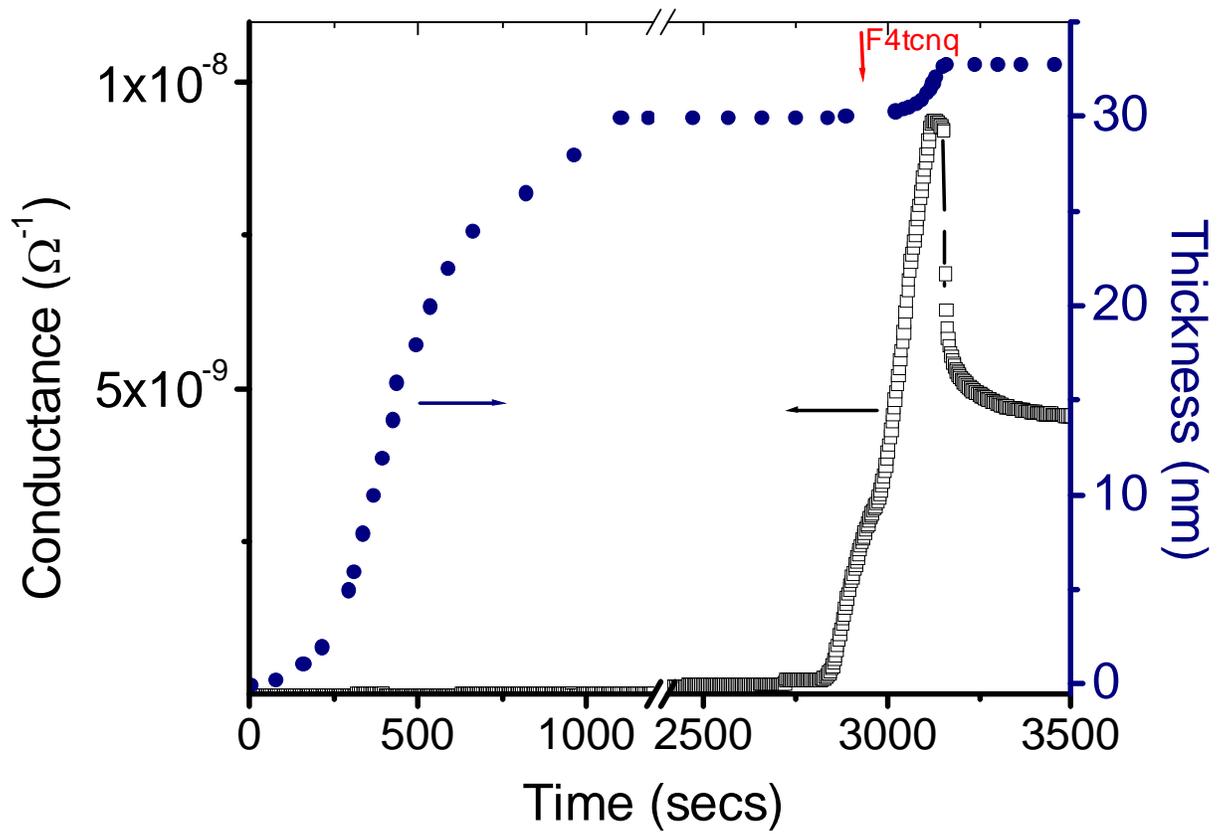

Figure 5

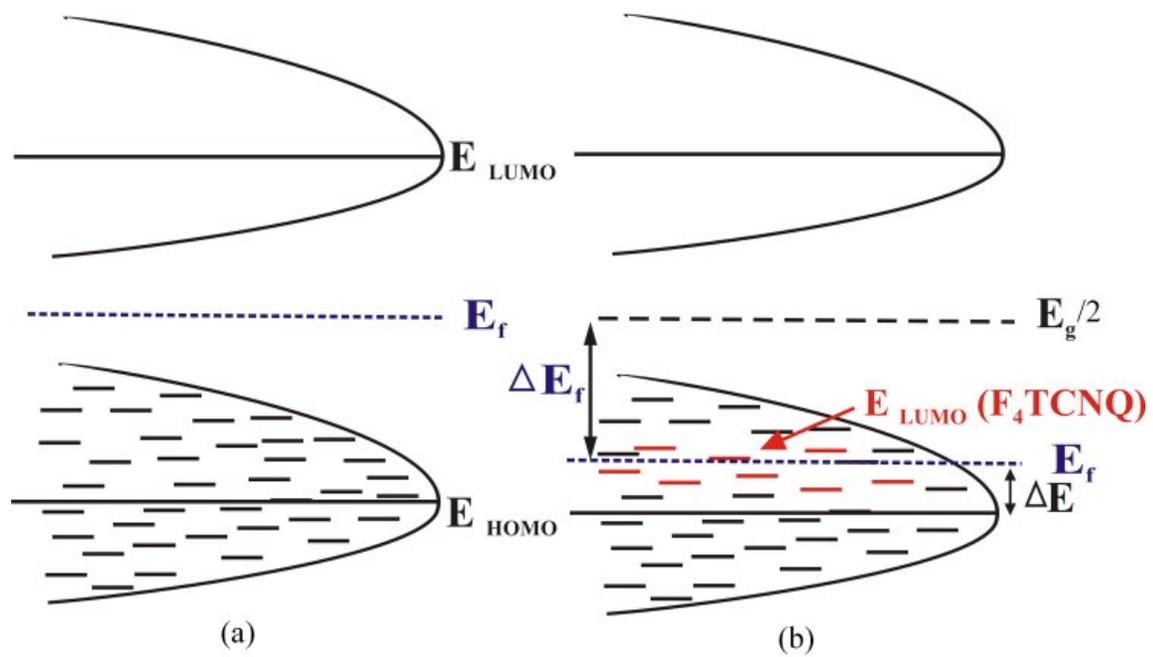

Figure 6

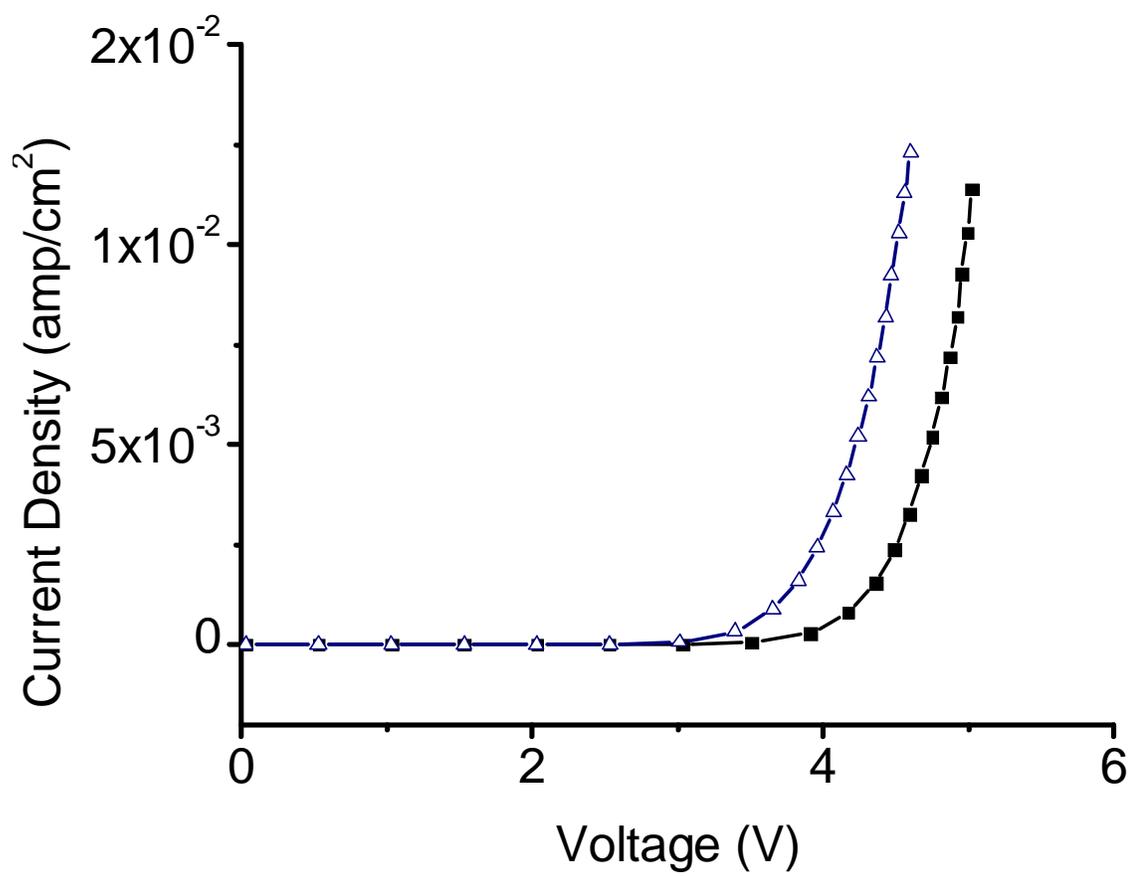